\documentstyle[preprint,aps]{revtex}

\begin{document}
\draft


\title{\bf Kolmogorov's law for two-dimensional 
electron-magnetohydrodynamic turbulence}

\author{A. Celani}
\address{Dipartimento di Ingegneria Aeronautica e Spaziale, 
Politecnico di Torino, \\ 
Corso Duca degli Abruzzi 24, 10129 Torino, Italy \\
Istituto Nazionale Fisica della Materia, Unit\`a di Torino}
\author{R. Prandi}
\address{Dipartimento di Fisica Teorica, Universit\`a di Torino, 
Via Pietro Giuria 1, 10125 Torino, Italy \\
Istituto Nazionale Fisica della Materia, Unit\`a di Pisa} 
\author{G. Boffetta}
\address{Dipartimento di Fisica Generale, Universit\`a di Torino, 
Via Pietro Giuria 1, 10125 Torino, Italy \\
Istituto Nazionale Fisica della Materia, Unit\`a di Torino}

\maketitle
\begin{abstract} The analogue of the Kolmogorov's four-fifths law 
is derived 
for two-dimensional, homogeneous, isotropic EMHD turbulence 
in the energy cascade inertial range.
Direct numerical simulations for the freely decaying case 
show that this relation holds true for different values
of the adimensional electron inertial length scale, $d_e$.
The energy spectrum is found to be close to the expected 
Kolmogorov spectrum.
\end{abstract}


The statistical theory of three-dimensional fully developed 
hydrodynamic turbulence relies on one outstanding issue:
the nonlinear transfer of energy from large to small scales
\cite{K41,Frisch,Batch53}.
It is therefore interesting to look for two-dimensional turbulent 
fluid dynamical systems sharing this same feature. Actually many 
of them exhibit a reversed energy flux, from the small scales to 
the larger ones, as is the case of 2D Navier-Stokes turbulence
\cite{McWill84,Brachet88,Legras88,Smi93},
Hasegawa-Mima turbulence \cite{Kukh95} or its geophysical 
counterpart, equivalent barotropic turbulence \cite{Larich90}. 

In this framework 2D  electron-magnetohydrodynamic (EMHD) turbulence 
deserves special attention, beyond its modeling applications,
since it has been shown to display, for the freely decaying case,
a forward energy cascade \'a la Richardson-Kolmogorov \cite{Bisk96}.

In this Letter it is introduced a relation which is the counterpart 
of the Kolmogorov four-fifths law for homogeneous and 
isotropic 2D EMHD turbulence. Its content is compared with the 
results obtained by direct numerical simulations.
 
EMHD equations are a fluid dynamical model for a cold electron 
plasma, moving in a uniform charge-neutralizing background of 
stationary ions.
In recent years this model has received considerable interest 
for its relation to inertially confined plasma and to laser-plasma 
interactions, but the comparison with experimental results is 
limited by the fact that plasma which evolve according to EMHD 
equations are usually short-lived. 
 
The equations for the electron plasma 
in adimensional form are \cite{King90} 
\begin{equation}
d_e^2 \displaystyle{{\partial {\bbox v} \over \partial t}} 
+ d_e^2 {\bbox v} \cdot \nabla {\bbox v} = 
-\nabla p - {\bbox E} -{\bbox v} \times {\bbox B}+ 
\nu_q (-\nabla)^{2q}{\bbox v} \; .
\label{eq:p}
\end{equation}
The velocity field ${\bbox v}$, the magnetic field ${\bbox B}$, 
and the electric field ${\bbox E}$ are allowed to have a nonzero 
component in any direction of three-dimensional space, meanwhile 
their functional dependence is restricted to the plane coordinates 
(${\partial \over \partial x_3} \equiv 0$).
The equations have been adimensionalized with respect to 
the characteristic macroscopic length $L$, the typical magnetic 
field $B_0$, the characteristic time $\tau = m_e c/(e B_0 d_e^2)$ 
and $d_e=[m_e c^2 /(4 \pi e^2 n L^2)]^{1/2}$ is the 
ratio of the inertial electron length scale to the 
integral scale $L$.
The density of the number of electrons $n$ is assumed to be uniform 
according to the incompressibility of the velocity field 
$\nabla \cdot {\bbox v} = 0$.
The approximation made by considering motionless ions requests
that the adimensional ion inertial length, $d_i$, must be larger 
than unity, thus limiting the range of admissible values 
for $d_e=d_i(m_e/m_i)^{1/2}$ to the interval 
$ d_e \stackrel{\sim}{>} 0.02$. 
As long as we are dealing with a fluid description of plasma, 
all the lengthscales under consideration must largely exceed 
the Larmor radius.
For stationary ions and negligible displacement current, the 
Amp\`ere law becomes
\begin{equation}
{\bbox v}= - \nabla \times {\bbox B} \; .
\label{eq:v}
\end{equation}
The last term in equation (\ref{eq:p}) 
is a dissipative term which mimics the 
effects of electron viscosity $(q=1)$ or resistivity $(q=0)$.
The total energy of the electron fluid
\begin{equation} 
E = \frac{1}{2}\int ( d_e^2 {\bbox v}^2 + {\bbox B}^2 ) d^2x
\label{eq:consE}
\end{equation}
is conserved by these equations in the ideal, non-collisional case. 

In the spirit of Kolmogorov analysis \cite{K41,Frisch}, one takes 
under consideration the spectral energy budget
\begin{equation}
\displaystyle{\partial \over \partial t} {\cal E}(K)+\Pi(K)
= -{\cal D}(K)
\label{eq:budg}
\end{equation}
where ${\cal E}(K)=\int_0^K E(k)\,dk$ is the mean cumulative energy
per unit mass contained at wavenumbers smaller than $K$, and 
$E(k)$ is the energy spectrum.
$\Pi(K)$ is the energy flux (per unit mass) from wavenumbers 
$k \leq K$ to larger wavenumbers, and ${\cal D}(K)$ is the 
cumulative energy dissipation up to wavenumber $K$.
Since the dissipation is localized to high wavenumbers 
$k \geq K_d$, there is a range of wavenumbers 
$K_0 \ll K \ll K_d$ where ${\cal D}(K)\simeq 0$, and the energy 
flux $\Pi(K)$ is determined by the inertial transfer of energy 
from the energy containing eddies at wavenumbers around $K_0$. 
When the large scale energy input due to the straining of
energy containing eddies at scale $K_0$ is equilibrated by 
dissipation taking place at small scales, one expects
the energy flux through wavenumber $K$, $\Pi(K)$, 
to be independent of $K$ \cite{Batch53}.
Actually it must be remarked that, due to the energy decay, the flux 
approaches a constant value, equal to the total dissipation,
only for very large $K$, and the crossover to the asymptotic 
behavior is very slow \cite{Borue95}.
In an analogous fashion as in three-dimensional hydrodynamic 
turbulence, in the limit of vanishing viscosity, 
$\nu \rightarrow 0$, the energy flux is expected to achieve a 
finite positive limit, depending on the value of $d_e$,
$\bar{\varepsilon} > 0$, 
\begin{equation}
\begin{array}{lcr}
\Pi(K) \simeq \bar{\varepsilon} &\hspace{20pt} 
& K_0 \ll K \ll K_d \; .
\end{array}
\label{hyp}
\end{equation}
This is a strong request which in 3D-hydrodynamics has 
experimental evidence; in the case under consideration it will 
be shown that numerical simulations
provide reasonable support to this hypothesis.

Assuming statistical homogeneity
the energy flux, $\Pi(K)$ can be expressed by means of 
physical space statistics by performing the Fourier transform of
(\ref{eq:budg}). The result is
\begin{equation}
\Pi(K)=\displaystyle{ {1 \over 2\pi} \int d^2\ell \; 
K {J_1(K \ell) \over \ell} \varepsilon({\bbox \ell}) } \; .
\label{eq:pi}
\end{equation}
where $J_1$ is the first-order Bessel function of the first kind,
and the energy flux in the physical space is given by
\begin{equation}
\varepsilon({\bbox \ell})=-\displaystyle{{\partial 
\over \partial t}}\displaystyle{ {1 \over 2}}
\langle d_e^2 {\bbox v}({\bbox x}) \cdot {\bbox v}({\bbox x}
+ {\bbox \ell}) + {\bbox B}({\bbox x}) \cdot {\bbox B}({\bbox x}
+ {\bbox \ell}) \rangle \big|_{NL}
\label{eq:eps}
\end{equation}
where the subscript $NL$ stands for the nonlinear contribution to 
the time derivative of the fields, as it can be extracted by the
equations of motion (\ref{eq:p}), and the brackets 
$\langle \ldots \rangle$ express ensemble averages. 
Using (\ref{eq:p}) and (\ref{eq:v}), making repeatedly use of 
statistical homogeneity, of incompressibility of the velocity 
field and solenoidality of the magnetic field, one obtains the 
following relation
\begin{equation}
\begin{array}{ll}
\varepsilon({\bbox \ell})= & 
\displaystyle{\partial \over \partial \ell_i} \Big(
\displaystyle{ - {d_e^2 \over 4}}
 \langle ( \delta {\bbox v} \cdot \delta {\bbox v}) 
\delta v_i \rangle + 
\vspace{2pt} \\   &
\displaystyle{ {1 \over 4}\langle (\delta {\bbox v} \cdot 
\delta {\bbox B}) \delta B_i \rangle } - 
 \displaystyle { {1 \over 8} \langle (\delta {\bbox B} \cdot 
\delta {\bbox B}) \delta v_i \rangle  }  \Big) 
\end{array}
\label{eq:eps2}
\end{equation}
with $i=1,2$\, denotes the planar components.
The expression (\ref{eq:eps2}) for the physical space energy flux
is the analogue of the K\'arm\'an-Howarth-Monin relation 
\cite{Frisch}, and it involves only differences of dynamical 
fields as $\delta {\bbox v} = {\bbox v}({\bbox x}
+{\bbox \ell})-{\bbox v}({\bbox x}).$ 
To proceed further one assumes statistical isotropy and
it is then possible to show that the physical space energy flux is
\begin{equation}     
\begin{array}{ll}
\varepsilon(\ell)= &
 \displaystyle{{1 \over 4}}(2+\ell \partial_{\ell}) \Big\{
-{d_e^2 \over 3}(4+\ell \partial_{\ell})
\displaystyle{{S_3(\ell) \over \ell}} - 
d_e^2  \displaystyle{{V_3(\ell) \over \ell}} +  \vspace{2pt} \\
 &
 \displaystyle{{1 \over 2}} (2+\ell \partial_{\ell}) 
\displaystyle{{T_3(\ell) \over \ell}}
 - \displaystyle{U_3(\ell) \over \ell}  
 + \displaystyle{W_3(\ell) \over \ell}  
 - {1 \over 2} \displaystyle{X_3(\ell) \over \ell}  
  \Big\}
\end{array}
\label{eq:epsfin}
\end{equation}
in which appear the following third order structure functions
\begin{equation}
\begin{array}{ll}
S_3(\ell)=\langle \delta v_{\parallel} 
\delta v_{\parallel} \delta v_{\parallel} \rangle \; ,& 
T_3(\ell)= \langle \delta v_{\parallel} \delta B_{\parallel} 
\delta B_{\parallel} \rangle \; ,  \vspace{4pt}\\
U_3(\ell)= \langle \delta v_{\parallel} \delta B_{\perp} 
\delta B_{\perp} \rangle \; , &
V_3(\ell)=\langle  \delta v_3 \delta v_3 \delta v_{\parallel} 
\rangle \vspace{4pt} \;,\\ 
W_3(\ell)= \langle \delta v_3 \delta B_3 \delta B_{\parallel} 
\rangle \; ,&
X_3(\ell)=\langle \delta B_3 \delta B_3 \delta v_{\parallel} 
\rangle \; ,
\end{array}
\end{equation}
where the standard notation for longitudinal, 
$\delta v_{\parallel}=\delta v_i \ell_i /\ell$,
and transverse differences, 
$\delta v_{\perp}=\epsilon_{ij}\delta v_i \ell_j /\ell$, 
has been used.

As a consequence of hypothesis (\ref{hyp}), it can be shown by a 
saddle-point argument that the physical space flux, 
$\varepsilon(\ell)$, must behave as  
\begin{equation}  
\begin{array}{lcr}
\varepsilon(\ell) \simeq \bar{\varepsilon} &\hspace{20pt} 
& \lambda \ll \ell \ll \ell_0 \;,
\end{array}
\label{hyp2}
\end{equation}
in the limit of vanishing viscosity, where the inertial range 
of lengthscales is now delimited by the ``Taylor scale'', 
$\lambda=(\nu_q E/D)^{1/2q}$, where $D$ is the energy dissipation,
and the energy containing scale $\ell_0 \sim 1/K_0$.
Inserting the expression for the energy flux (\ref{eq:epsfin}) 
inside relation (\ref{hyp2}), one obtains the 2D EMHD counterpart 
of Kolmogorov's four-fifths law \cite{K41,Frisch}
\begin{equation}
Q_3(\ell) \simeq \bar{\varepsilon}\ell
\label{eq:str}
\end{equation}
where
\begin{equation}
\begin{array}{ll}
Q_3(\ell) = & \displaystyle{
 -{2 \over 3} d_e^2 \, S_3(\ell)-{1 \over 2} d_e^2 \, V_3(\ell) + 
{1 \over 2} T_3(\ell)  } \vspace{3pt}  \\ & \displaystyle{
-{1 \over 2} U_3(\ell) +{1 \over 2} W_3(\ell)
-{1 \over 4} X_3(\ell) } \; .
\end{array}
\label{eq:str1}
\end{equation}
Relation (\ref{eq:str}) relies only on the aforementioned 
hypothesis here recalled: homogeneity, isotropy, and
the existence of an inertial range of wavenumbers in which the 
energy flux is constant, with a value
tending to a finite positive limit for vanishing viscosity. 
The most remarkable aspect of relation (\ref{eq:str},\ref{eq:str1}) 
is that it does not only provide a linear scaling for the third 
order structure function $Q_3(\ell)$ within the inertial range of 
lengthscales, but it also prescribes the value of the numerical 
coefficient appearing in front of the scaling relation.
Moreover, it is valid for any value of $d_e$, meanwhile no
power law scaling relation is expected to hold for, say, the 
second order structure functions, apart from limiting cases such 
as $d_e \sim 1$ and $d_e \ll 1$ \cite{Bisk96}.

To check the validity of EMHD Kolmogorov law (\ref{eq:str}),
the equations (\ref{eq:p}) are solved in a square box of size 
$2 \pi \times 2 \pi$ imposing periodic boundary conditions, by means 
of a standard pseudospectral method with resolution $N \times N$.
Hyperdissipation with $q=4$ is employed in order to achieve a 
larger extent of the inertial range, which in physical space is 
known to be much narrower than in spectral space \cite{Frisch}. 
Hyperviscosity is set to $\nu_4=10^{-12},10^{-13}$ for $N=512$ 
simulation and $\nu_4=10^{-13}$ for $N=1024$.

The initial conditions are 
$ v(k) \sim k^2 \exp(-k^2/2k_0^2)$ with random phases, $k_0=1$ 
and total energy of order unity in both resolution simulations.
After a transient of a few large eddy turnover times, when the 
energy initially contained at the lowest wavenumbers starts to
cascade down to small scales, the energy dissipation reaches a 
maximum value and then a self similar stage of decay sets in 
\cite{Bisk96}.
The energy flux is approximately constant throughout the inertial 
range of wavenumbers (see Fig.1), and, in agreement with the 
assumption (\ref{hyp}), its value appears to be asymptotically 
independent of viscosity.
The structure function $Q_3(\ell)$ is computed during the self 
similar stage of decay. The results are obtained after averaging 
over a short time in order to get better statistics at small scales.
As shown in Fig.2 the compensated structure function
$Q_3(\ell)/(\bar{\varepsilon}\ell)$ approaches unity, as prescribed 
by the relation (\ref{eq:str}) in an interval 
delimited from below by the ``Taylor scale'' $\lambda$
and above by the energy containing scale $\ell_0$.
By lowering the viscosity (crosses, $N=1024$) the scaling range 
extends to smaller scales over almost one decade. 
As previously remarked, the width of the inertial range is actually
diminished by the fact that, for a decaying flow, the energy flux in
wavenumber space is not constant except asymptotically (see Fig.1).
This results are an evident numerical confirmation of
the validity of the Kolmogorov-type relation 
(\ref{eq:str}). This kind of assessment is important since 
it lies at the foundations of the statistical study of turbulence.

Introducing the further hypothesis of statistical self-similarity
one can infer the following scaling behavior for the velocity and 
for the magnetic field differences in the asymptotic case
\begin{equation}
\begin{array}{l}
\delta v (\ell) \propto \bar{\varepsilon}^{\,1/3} \ell^{\,1/3} 
\vspace{2pt} \\
\delta B (\ell) \propto \bar{\varepsilon}^{\,1/3} \ell^{\,4/3} 
\end{array}
\hspace{30pt}
\begin{array}{r}
\ell \ll d_e \; ,
\end{array}
\label{eq:scal}
\end{equation}
which leads to the small scales energy spectrum, 
dominated by kinetic energy, 
\begin{equation}
\begin{array}{lcr}
E(k) = C_K \bar{\varepsilon}^{\,2/3}
k^{-5/3} & \hspace{30pt} & kd_e \gg 1 \;.
\end{array}
\label{eq:spec}
\end{equation}
On the other hand, for scales larger than $d_e$, the expected 
self-similar scaling  is
\begin{equation}
\begin{array}{l}
\delta v (\ell) \propto \bar{\varepsilon}^{\,1/3}\ell^{-1/3} 
\vspace{2pt} \\
\delta B (\ell) \propto \bar{\varepsilon}^{\,1/3}\ell^{\,2/3}
\end{array}
\hspace{30pt}
\begin{array}{r}
\ell \gg d_e
\end{array}
\label{eq:scal2}
\end{equation}
leading to the energy spectrum, dominated by magnetic energy,
\begin{equation}
\begin{array}{lcr}
E(k)=C'_K \bar{\varepsilon}^{\,2/3}
k^{-7/3} & \hspace{30pt} & kd_e \ll 1 \; .
\end{array}
\label{eq:spec2}
\end{equation}
The slopes of the computed spectra, as shown in Fig.3, are close 
to the estimates (\ref{eq:spec}) and (\ref{eq:spec2}) 
over a wide range of wavenumbers.

The main results of this work are the derivation
of a Kolmogorov-type relation for 2D EMHD decaying turbulence,
and its numerical confirmation.
Since the Kolmogorov's law stands as a corner stone in the study 
of the statistical features of turbulence, these results  
form the basis of further analysis, starting from the issue of 
intermittency. 
 
The authors wish to acknowledge support and hospitality
by the Istituto di Cosmogeofisica CNR, Torino.  
The calculations were partially performed with computer facilities
of INFN, Sezione di Torino.
                 

\begin{figure}

\caption{Energy flux in wavenumber space normalized to energy
dissipation.
(a) $d_e=0.3$. Continuous line: $N=512$; $\nu_4=10^{-12}$.
Dotted line: $N=1024$; $\nu_4=10^{-13}$.
(b) $d_e=0.02$. Continuous line: $N=512$; $\nu_4=10^{-13}$.}
\end{figure}

\begin{figure}

\caption{Compensated structure function 
$Q_3(\ell)/(\bar{\varepsilon}\ell)$.
(a) $d_e=0.3$; Diamonds: $N=512$; $\nu_4=10^{-12}$.
Crosses: $N=1024$; $\nu_4=10^{-13}$.
(b) $d_e=0.02$; Diamonds: $N=512$; $\nu_4=10^{-13}$.}
\end{figure}

\begin{figure}

\caption{Energy spectrum.
(a) $d_e=0.3$; Diamonds: $N=512$; $\nu_4=10^{-12}$.
Crosses: $N=1024$; $\nu_4=10^{-13}$.
The continuous line is the Kolmogorov spectrum 
(\protect\ref{eq:spec}) with $C_K=2.0$.
(b) $d_e=0.02$; Diamonds: $N=512$; $\nu_4=10^{-13}$.
The continuous line is the spectrum (\protect\ref{eq:spec2})
with $C'_K=8.0$}

\end{figure}


\begin{thebibliography}{99}

\bibitem{K41} A.N. Kolmogorov, Dokl. Akad. Nauk SSSR {\bf 32}, 
16 (1941)

\bibitem{Frisch} U. Frisch, {\it Turbulence. The legacy of 
A.N. Kolmogorov \/}
(Cambridge University Press, Cambridge, 1995) 

\bibitem{Batch53} G. K. Batchelor, {\it The theory of homogeneous 
turbulence \/} (Cambridge University Press, Cambridge, 1953)

\bibitem{McWill84} J. C. McWilliams, J. Fluid Mech. {\bf 146}, 
21 (1984).

\bibitem{Brachet88} M. Brachet, M.Meneguzzi, H. Politano, and 
P. Sulem, J. Fluid Mech. {\bf 194}, 33 (1988)

\bibitem{Legras88} B. Legras, P. Santangelo, and R. Benzi, 
Europhys. Lett. {\bf 5}, 37 (1988)

\bibitem{Smi93} L. M. Smith and V. Yakhot, Phys. Rev. Lett.
{\bf 71}, 352 (1993)

\bibitem{Kukh95} N. Kukharkin, S. A. Orszag, and V. Yakhot, 
Phys. Rev. Lett. {\bf 75}, 2486 (1995)

\bibitem{Larich90} V. D. Larichev and J. C. McWilliams, 
Phys. Fluids A {\bf 3}, 938 (1991)

%
%
%
%

\bibitem{Bisk96} D. Biskamp, E. Schwarz, and J. F. Drake, 
Phys. Rev. Lett {\bf 76}, 1264 (1996).

\bibitem{King90} A. S. Kingsep, K. V. Chukbar, and V. V. Yan'kov, 
in {\it Reviews of Plasma Physics \/} 
(Consultants Bureau, New York, 1990), Vol.16, 243.

\bibitem{Borue95} V. Borue and S. A. Orszag, 
Phys. Rev. E {\bf 51}, 857 (1995)

\end{thebibliography}
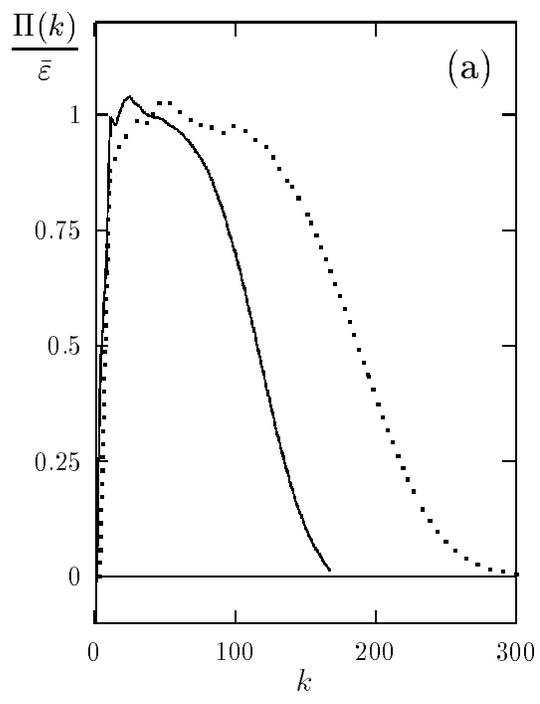
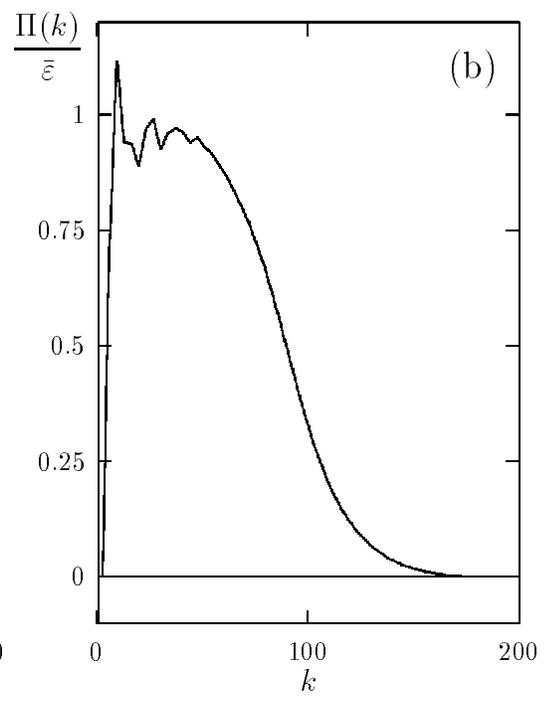
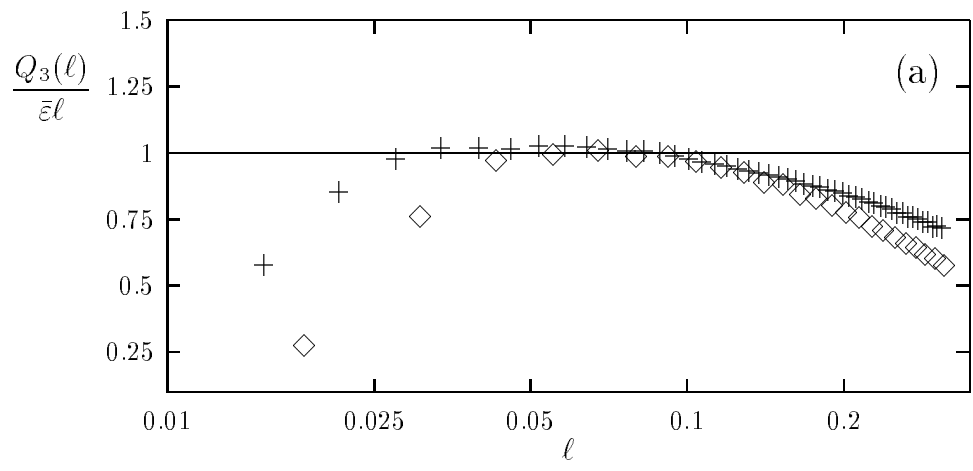
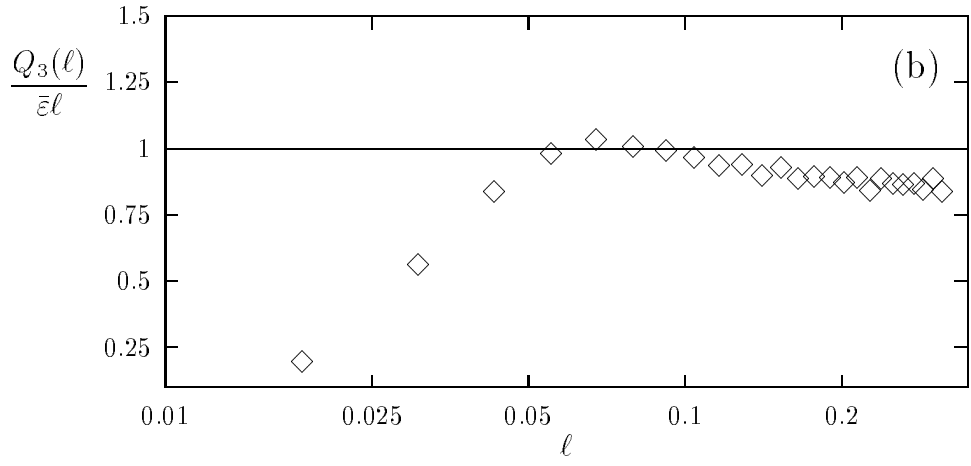
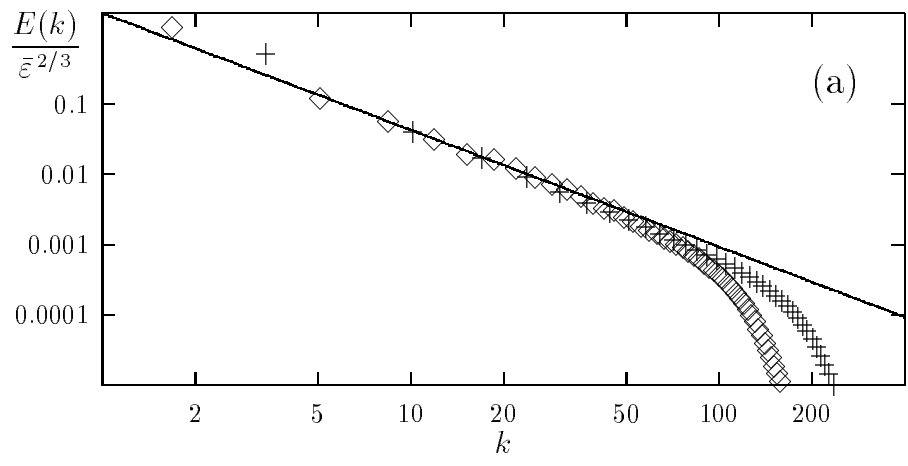
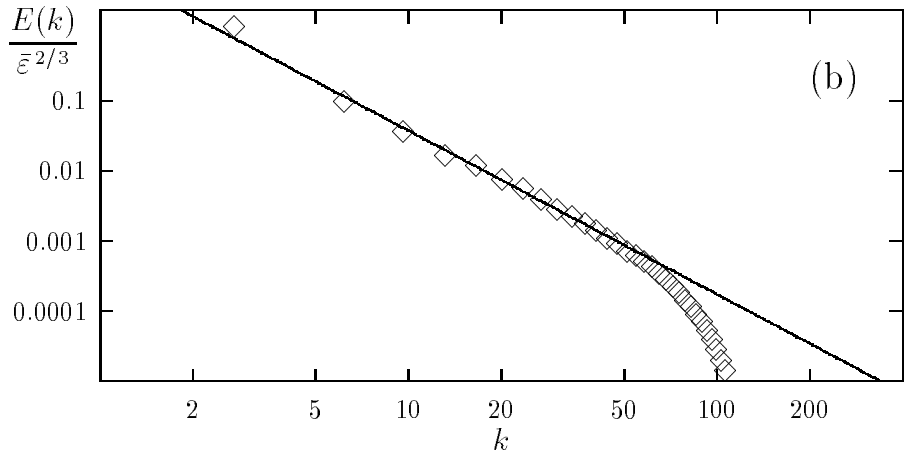
\end{document}